\documentclass[article]{elsarticle}

\usepackage {color}

\usepackage{lineno,hyperref}
\modulolinenumbers[5]

\usepackage{graphicx}
\usepackage{amsmath}
\usepackage{enumerate}
\usepackage{tikz}
\usepackage{subcaption}
\usepackage{svg}

\usepackage{natbib}

\journal{Journal}

\begin{document}

\begin{frontmatter}
			
\title{``Water to the ropes'': a predictive model for the supercontraction stress of spider silks}
\author{Vincenzo Fazio$^1$,  Nicola Maria Pugno$^{1,2*}$,\\ Giuseppe Puglisi$^{3**}$.}

\address{$^{1}$ Laboratory for Bioinspired, Bionic, Nano, Meta Materials \& Mechanics,\\ University of Trento, Via Mesiano 77, 38123 Trento, Italy;}
\address{$^{2}$ 
	School of Engineering and Materials Science, Queen Mary University of London,\\ Mile End Road, London E1 4NS, U.K.;}
\address{$^{3}$ Department of Civil Environmental Land Building Engineering and Chemistry, Polytechnic University of Bari, via Orabona 4, 70125 Bari, Italy.}

\address{$^{*}${nicola.pugno@unitn.it}}
\address{$^{**}${giuseppe.puglisi@poliba.it}}

\begin{abstract}
When humidified at different moisture conditions, restrained spider silk fibers can exhibit a very high supercontraction phenomenon. The hydration water molecules induce a Hydrogen-bonds disruption process that, due to entropic effects, decreases the natural -zero force- end-to-end chains length. By considering a bundle of macromolecules, we describe supercontraction as a possible actuation system and determine the maximum actuation force depending on the silk properties at the molecular scale and on the constraining system representing other silk threads or the actuated device. The  comparison with experimental results of {\it Argiope trifasciata} silk fibers show the effectiveness of the proposed model in quantitatively predicting the experimental actuation properties. The considered historical case study of obelisk rescue in Saint Peter’s Square (Rome)  through ropes hydration is discussed evidencing the optimal performances of this natural material adopted as  moisture powered actuator: we obtain a  work density of 2.19 kJ/m$^3$ making spider silk the most performant hydration driven active material. Moreover we obtain a power density of the order of 730 W/kg about three times the most performant carbon nanotube actuators making such material very competitive as compared with all types of actuator. The analytic description of the macroscopic actuation parameters from microscale properties shows the possibility of adopting our approach also in the field of bioinspired artificial silks design, possibly considering also important non-linear effects in the actuated system. 
\end{abstract}

\end{frontmatter}

\section{Introduction}
	Spider silks have been increasingly the focus in very wide research and technological fields due to their extreme mechanical properties~\cite{gosline1986structure} such as extraordinary strength and toughness, self-healing, and environmental adaptability. Such properties are often unattained by artificial materials (see \cite{perez2021basic} and references therein). This outstanding material response, resulting by a complex hierachical  structure organization, attracted the attention in the important field of biomimetics \cite{zhao2014bioinspired,greco2021tyrosine,arndt2022engineered}. The deduction of the fiber material response, starting from silk structure at the molecular scale, represents a demanding theoretical problem, not completely clear, especially regarding the adaptability to different environmental (humidity and temperature) and loading conditions. 
	
	Here, by extending the recent results in \cite{FDPP}, we focus on the important role of hydration on the actuation material properties of spider silks. Indeed, a striking effect observed in spider silks is the so called {\it supercontraction effect}, addressed, to the knowledge of the author, for the first time in 1977~\cite{Work_1977}, that occurs when a spider silk thread is exposed to humidity.
	Depending on the silk composition, the experiments show the existence of a Relative Humidity (RH) threshold beyond which the fiber contracts up to a half of its initial (dry) length. The experimentally observed contraction depends on several factors, including spider species~\cite{boutry2010evolution}, type of silk (among the up seven different ones that some spiders can produce \cite{vollrath1992spider,gosline1994elastomeric}), environmental conditions \cite{plaza2006thermo} and hydration rate \cite{agnarsson2009supercontraction}. 
	
Interestingly, thinking both to the humidity effects on spider webs and to the possibility of adopting such behavior as a natural actuation device, constrained humidified silks generates a stress that can be measured trough a load cell \cite{Guinea2003,ene2011supercontraction}. In the case of fully locked end-points (fixed end-to-end length) the higher magnitude of the stress is attained, typically of the order of tens of MPa  as experimentally measured  by \cite{agnarsson2009supercontraction,Guinea2003,blackledge2009super}. 
Clarifying and predicting the mechanisms that originate this phenomenon, both in natural and artificial silks, is considered of great interest also in the perspective of adopting this behavior at the base of mechanical actuation  \cite{dong2021programmable} or humidity sensing devices \cite{zhang2022spider}. This aspect is the focus of the following analysis.

	Two types of experiments are typically performed to describe the supercontraction effect. In the first type, the dry silk sample is free to shorten and the length variation under increasing humidity is measured. In the second case the fiber is fixed at its dry length and no contractions is allowed and the force applied to fix the length at increasing humidity is measured. The maximum attained value of the stress is then measured, {\it supercontraction stress} \cite{bell2002supercontraction,savage2004supercontraction} and it represents the previously recalled actuation force. Intermediate boundary conditions (see \cite{Florio:2019wc} for a theoretical discussion of the influence of different boundary conditions on the material response of constrained systems) represent real phenomena with the silk constrained by deformable devices such as other silks fibers in the web  or external attaching systems. 

At the molecular scale, the supercontraction effect is due to hydration-induced hydrogen bonds breaking. More in detail, at the molecular scale spider silk is composed by an amorphous matrix of oligopeptide chains and pseudo-crystalline regions, made up principally of polyalanine $\beta$-sheets \cite{elices_hidden_2011,sponner_composition_2007} with dimensions between $1$ and $10$ nm \cite{keten_nanostructure_2010} oriented in the direction of the fiber \cite{jenkins_characterizing_2013}. The chains are highly hydrogen bonded, with a medium-low density of H-bonds in the amorphous part whereas the nanocrystallites are characterized by a high density of H-bonds \cite{yarger_uncovering_2018}.
Supercontraction is the result of the entropic recoiling of the macromolecules when the hydration water molecules break the H-bonds naturally present in the virgin silk, that previously fixed the macromolecules in a natural elongated conformation induced during the spinning of the fiber \cite{du2006design,elices2005finding,elices2011hidden}. 


In the following, we propose an approach to determine the supercontraction stress arising when a dry silk fiber is humidified with monotonically increasing {\small RH} and the length is fixed. 
Specifically, we reframe the multiscale approach for the humidity and thermomechanical response of spider silks effects recently proposed in \cite{FDPP}, where  the silk thread is modeled as a composite material made by hard crystalline fraction and a soft amorphous fraction embedded in an elastic tridimensional matrix reproducing the network effect \cite{flory_1982}. Regarding the hydration phenomenon, it is important to remark that, due to their different conformations and chemical composition, hard crystalline domains are hydrophobic whereas the soft amorphous fractions are hydrophilic. This results in two different responses to the hydration for the two different fractions. Indeed, the water molecules cannot penetrate the hydrophobic crystalline domains and the only humidity effect in this fraction is an increased misalignment of the crystallites with respect to the fiber axis. 

\begin{figure}[htb] \centering 	\includegraphics[width=\textwidth]{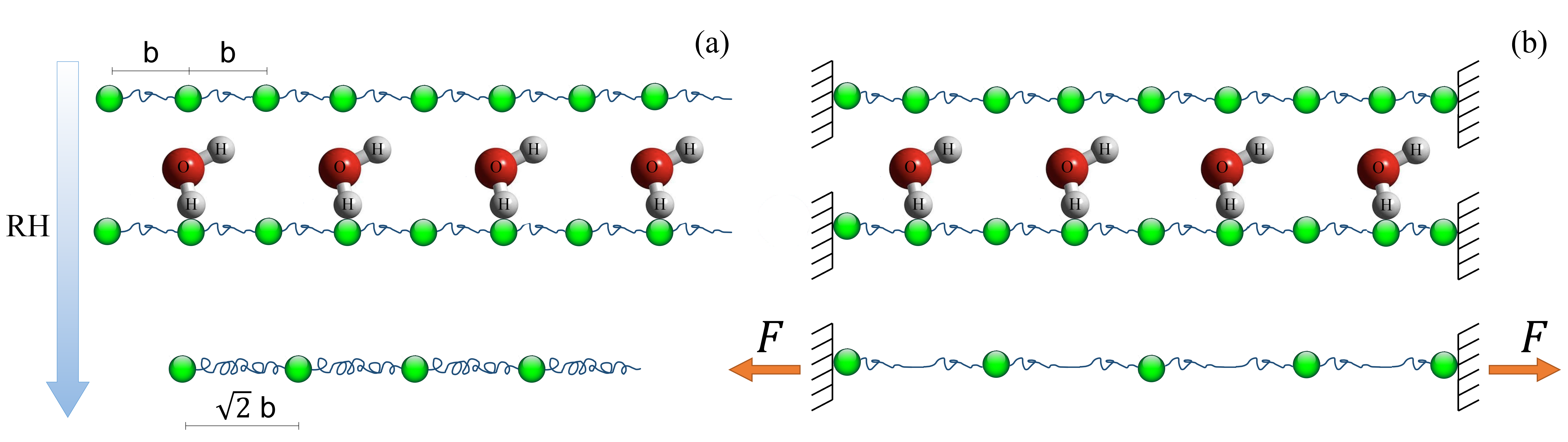}
	\caption{Cartoon of the supercontraction effect. Water molecules reduces the number of $H$ bonds and, due to entropic effects, the chain is subjected to a reduction of its natural end-to-end length. In (a) we show the case of humidification for an unconstrained chain and the resulting contraction: {\it e.g.} two Kuhn segments of total length $2b$ coalesce in a segment of length $\sqrt{2} b$. In (b) we show the case of (perfectly) constrained molecule and the resulting force applied to the constraint of particular interest in this paper where  actuation properties are considered.} \label{fig:unres_res}  \end{figure}

On the other hand, the hydration water molecules decrease the percentage of crosslinks in the softer region, inducing a modification of the natural (zero-force) length of the chains composing this region. We then assume that during hydration the natural lengths of the hard region and the matrix do not vary, so that the supercontraction stress is a direct effect of the variation of the H-bonds percentage and resulting variation of the natural length of the silk molecules belonging to the soft region (as detailed in \cite{FDPP}). As a consequence, in the specific experiment of fixed total length we may deduce that the mechanical response is due to the only configurational changes of the soft fraction (see the scheme in Fig.~\ref{fig:unres_res}(a)). 
Instead, as anticipated above, when restrained to its natural (zero force) length, if humidified the fiber cannot contract and supercontraction forces develop (see Fig.~\ref{fig:unres_res}(b)).

In this work we provide a model to predict the supercontraction stress when the environmental humidity is monotonically increased. Further, we enhance the previous model in \cite{FDPP} of the amorphous chain by reducing the number of fit parameters based on classical results of reaction kinetic theory. The effectiveness of the modifications of the model is demonstrated by quantitatively predicting  the experimental supercontraction stress for restrained spider Argiope trifasciata silk fibers.

We then show the efficiency of the actuation properties of spider silks fibers under external variable humidity conditions by considering a historical example when in 1586 Pope Sisto V  asked for the erection of the 350 tons and 25 meters high obelisk in Saint Peter’s Square (Rome, Italy). In that case the abrupt fall of the obelisk during erection was avoided by their activation shortening through hydration of the ropes adopted for the erection. To explicitly describe through a virtual example the incredible actuation properties of this material
we design a virtual system based on the actuation of spider silks to attain the same results of the ropes.

We then study the fundamental effect of the elasticity of the actuated device, typically neglected in previous studies, and show how actuation properties such as the actuation work and power depend not  only on the spider silks, but also on the stiffness of the interacting actuated device.
We then consider the optimization problem for a linear actuated systems to determine the  values of the relative (actuation and actuated) stiffness leading to the maximum actuation work. As a result, we are able to explicitly determine the actuation properties of silks and we show that a work density of 2.19 kJ/m$^3$ can be attained  so that spider silk, to the knowledge of the authors, can be considered the most performant hydration driven active material. Moreover we determine a power density of 730 W/kg that is about three times the most performant carbon nanotube actuators. We then deduce that also in a comparison with general actuation systems spider silks result as very competitive actuators. Eventually, as a simple prototypical amplified actuation device, we also show the possibility of strongly increasing the actuation by considering multistable responses of the actuated system. 

We believe that our study can open up the understanding and the application of spider silks and artificial silks as actuation devices. Based on our analytic results we are also confident that the proposed model will be important in the design of very efficient actuation materials and devices.

\section{Model}

Here for the spider thread we adopt the theoretical model  recently proposed in \cite{FDPP} where the authors describe the spider thread as composed by a multiphase material.
Thus the stress is additively decomposed as the sum of the the soft fraction contribution $\sigma_{\text{soft}}$, depending on {\small RH}, the hard fraction contribution $\sigma_{\text{hard}}$, active only for lengths larger than the natural one (see \cite{FDPP} for details), and the elastic matrix contribution, representing an external network, here modeled as Neo-Hookean with modulus $\mu$. 
The total constitutive response of the silk is then
\begin{equation}\sigma(\lambda)=\sigma_{soft}(\lambda)+\sigma_{hard}(\lambda)+\mu \left( \lambda_{\text{eq}}- \frac{1}{\lambda_{\text{eq}}^2}\right).\end{equation}
Notice that in this paper, the hard region chains are slack (they do not sustain any compressive load), so that here we may neglect their contribution. The extension to the contemporary application of external humidity and mechanical loading can be obtained by simply considering the contribution $\sigma_{\text{hard}}\neq 0$. 

On the other hand here we consider the case of fixed length and determine the resulting stress
$$\sigma_{act}=\sigma(1)\equiv \sigma_{soft}(1).$$

Let us then consider the only soft fraction. The silk thread (hygro-thermo-mechanical) behavior results as an average response of the (parallel) silks molecules spider silk \cite{plaza2006thermo}. Specifically, we suppose that the soft fraction is composed of identical molecules composed by a number $n$ of Kuhn segments each of length $b$.
The expectation value of the unloaded end-to-end length of an ideal chain can then be evaluated by using a classical result of the Statistical Mechanics \cite{rubinstein2003polymer} as
\begin{equation} \label{sm}
	L_n=<r^2>^{1/2}= n^{1/2} \, b, \end{equation}
whereas its contour length is $L_c= n \, b$.

\begin{figure}[h] \centering 	\includegraphics[width=.62\textwidth]{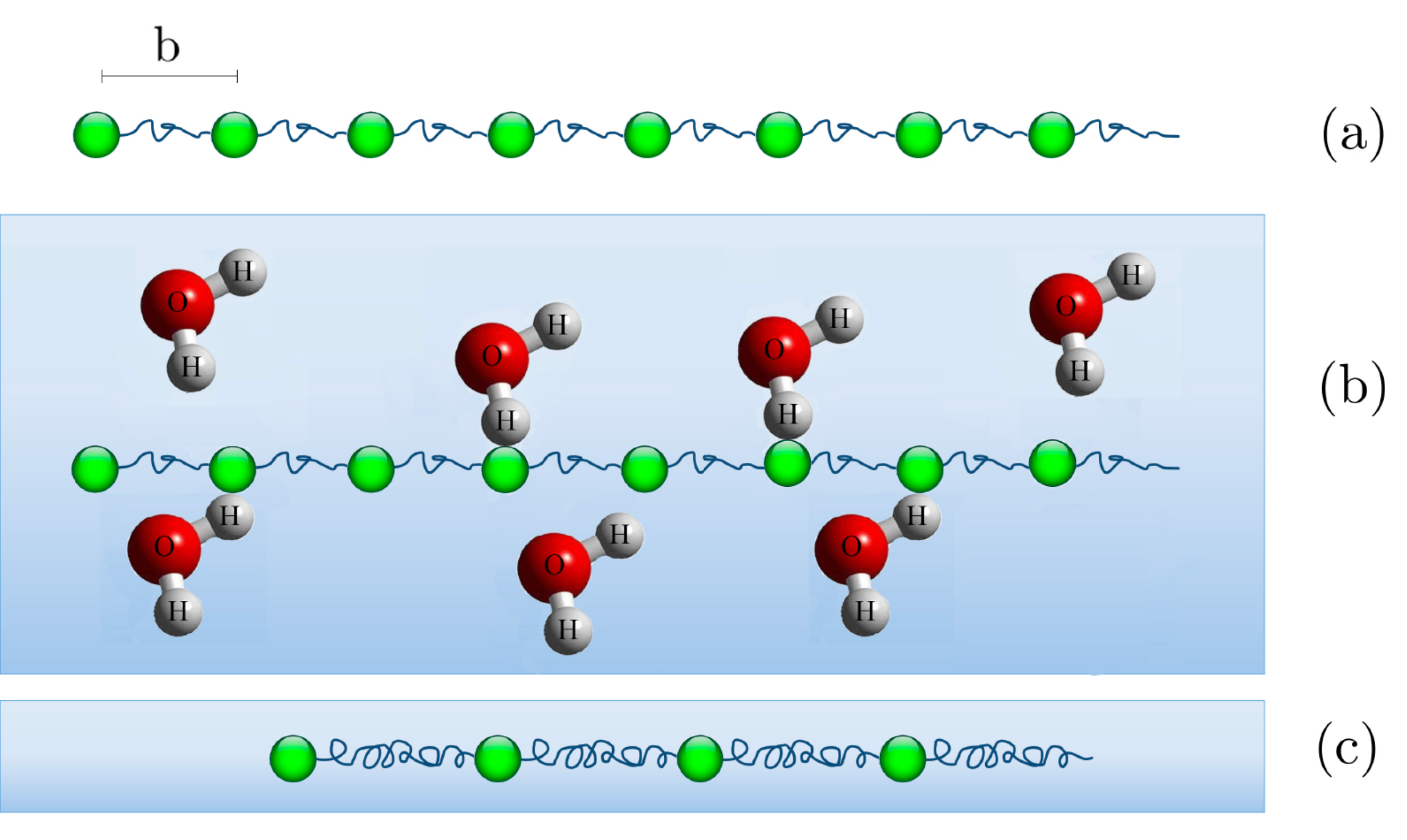}
	\caption{Scheme of the hydration reaction. An example of a chain composed by $m_0=8$ (blue) Kuhn segments of length $b$ in the initial dry condition is schematized in (a). In the example of wet environment shown in (b) we suppose to have $n_w=7$ water molecules and $m_b = 4$ water molecules that bound and break links, so that only $m=4$ links are kept in the wet condition (c).} \label{fig:scheme_links}  \end{figure}

We indicate by $m_b$ and $m$ the numbers of broken and unbroken H-bonds, respectively, with 
\begin{equation}\label{mo}m_b+m=m_o,\end{equation} 
where  $m_o$ is the initial (dry) number of bonds. We then describe the disruption process induced by the hydration water molecules $n_w$ \cite{du2006design} based on the classical Michaelis-Menten kinetics adopted to describe the enzymatic reaction as regulated by the concentration of a substrate. The considered reaction (see Fig.~\ref{fig:scheme_links}) is 
$$ m+n_w \underset{k^-}{\overset{k^+}\rightleftharpoons} m_b.$$
so that the rate of bonds breaking is
\begin{equation} \frac{\text{d} \, m_b}{\text{d}t}=k^+n_w(\text{\small RH}) \, m -k^-m_b. \end{equation}
When the equilibrium of the transition is attained, we get $\frac{m_b}{n_w(\text{\footnotesize RH}) \,m}=\frac{1}{k_d}$ with  $k_d=\frac{k^-}{k^+}$.
By substituting in  Eqn.~\eqref{mo} we obtain the Michaelis-Menten equation \cite{Johnson:2011uy}
\begin{equation} \label{mm} \frac{m(\text{\small RH})}{m_o}=\displaystyle \frac{1}{1+\displaystyle\frac{n_w(\text{\small RH})}{k_d}}. \end{equation}

\begin{figure}[htb] \centering 	
\includegraphics[width=.45\textwidth]{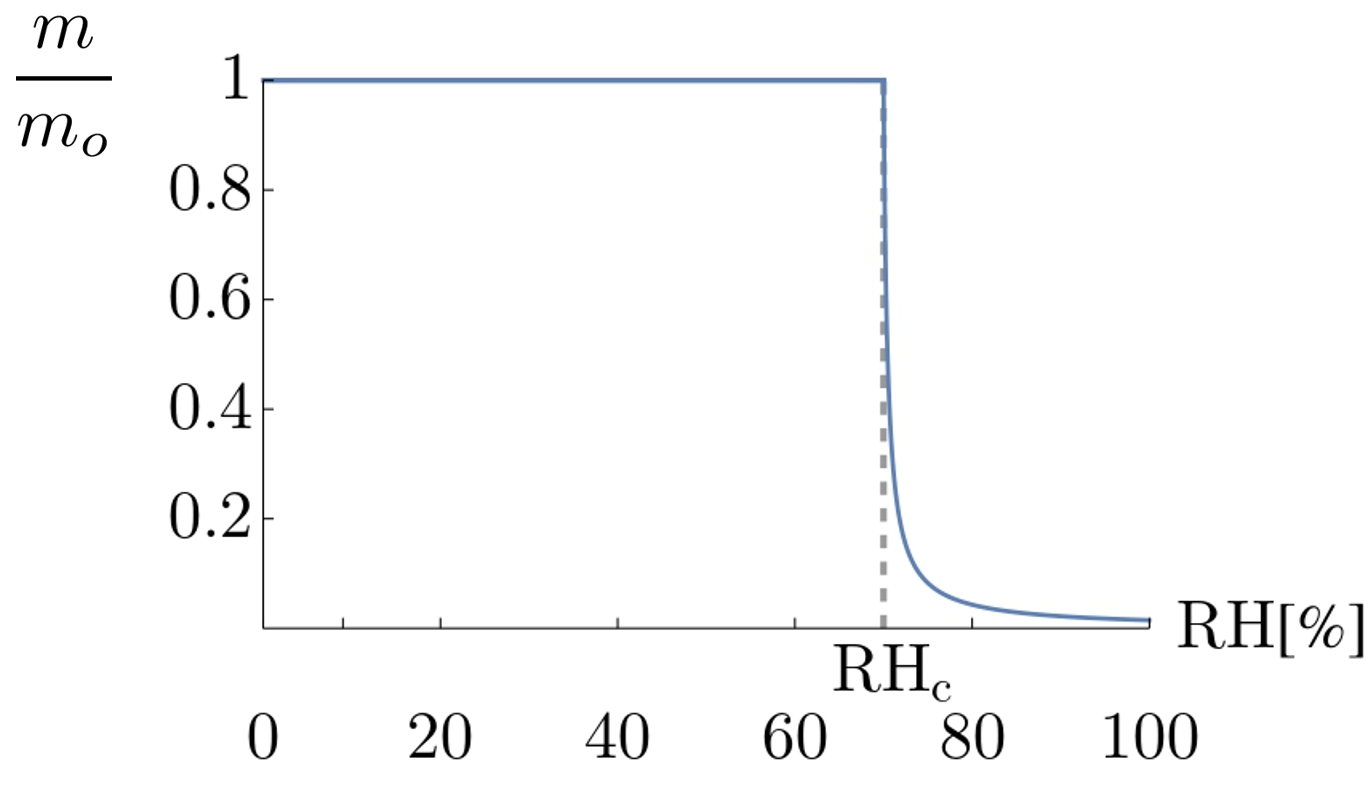} 	\caption{Influence of the humidity on the relative (as compared with the initial, dry condition) number of H-bonds. Assumed parameters {\small RH}$_c = 70\%,k=2.22$.
} \label{fig:rh_m}  \end{figure}

To take into account the experimental observation of the existence of a supercontraction threshold $\text{\small RH}_c$, such that as the relative humidity grows the spider silk fiber sharply contracts with increasing humidity as the threshold is approached from above {\small RH} \cite{fu2009moisture},  we consider the simple law
\begin{equation}n_w(\text{\small RH})
= 
	\begin{cases}
		0 & \text{ if \small RH}<\text{\small RH}_c,\\
		k_p(\text{\small RH}-\text{\small RH}_c) & \text{ if \small RH} \ge \text{\small RH}_c.
\end{cases}  \label{nwRH}
 \end{equation}Thus we get the final relation
\begin{equation} \label{mmo}
	\frac{m (\text{\small RH})}{m_o} = 
	\begin{cases}
		1 & \text{ if \small RH}<\text{\small RH}_c,\\
		\frac{1}{1+k(\text{\small RH}-\text{\small RH}_c) } & \text{ if \small RH} \ge \text{\small RH}_c,
\end{cases} \end{equation}
with 
\begin{equation}k=\frac{k_p\, k^+}{k^-}\label{kkk}\end{equation}
representing the main parameter regulating the hydration induced debonding effect of H-bonds.
The resulting trend is schematized in Fig.~\ref{fig:rh_m}.


Following \cite{FDPP}, to obtain the variation of the natural length of the `representative' chain with the humidity, we assign the initial (mean) number of chain monomers $n_o$ when the silk is spun (virgin) and we identify the number of links $m$ with the number of domains in which the chain is divided (see the scheme in Fig.~\ref{fig:scheme_links}). In the generic humidity state, the (mean) number of monomers in each domain is therefore $n(\text{\small RH})=n_o/m(\text{\small RH})$. Based on Eqn.~\eqref{sm} we determine the natural length of the whole chain:
\begin{equation} \label{ln}
	L_n(\text{\small RH})=m\sqrt{n_o/m(\text{\small RH})} \ b =\sqrt{m(\text{\small RH}) \ n_o} \ b.\end{equation}

Following \cite{de_tommasi_energetic_2013}, we adopt a Worm Like Chain (WLC) type energy density per unit chain contour length $L_c$, $\varphi_e=\varphi_e(L,L_c)=\kappa\frac{L^2}{L_c-L}$ where $\kappa=\frac{k_B T}{4 l_p}$, $T$ is the temperature, $k_B$ the Boltzmann constant, and $l_p$ the persistence length. 
This energy respects the limit extensibility condition, $\lim_{L \rightarrow L_c} \varphi_e(L,L_c)=+\infty$, and allows for explicit calculations.
Moreover, as described above, we consider that the end-to-end distance $L$ can be decomposed into a variable (zero-force) natural length measured by \eqref{sm} and the remaining elastic component $L_e= L-L_n$, as firstly proposed in \cite{jmbbm2021}. Thus we assume an energy and a force-elongation law for a single chain
\begin{equation} \label{eqn:force-stretch}\begin{array}{lll}
		\varphi_e&=&\kappa\frac{L_e^2}{L_c-L}\\
		f&=&\frac{\partial\varphi_e}{\partial L}=
		\kappa \left[\left( \frac{L_c-L_n}{L_c-L}\right)^2-1 \right], \end{array}\end{equation}
		
\noindent with the force decreasing to zero as the length attains its natural length ($L=L_n$ or $L_e=0$).


In order to deduce the macroscopic behavior of the thread, we consider the classical \textit{affinity hypothesis} \cite{rubinstein2003polymer} that identifies the macroscopic stretches with the macromolecular ones. 
We can then introduce the following stretch measures 
\begin{equation} \label{eqn:stretch measures}
	\begin{tabular}{ll}
		\hspace{.2cm}$\lambda=\frac{L}{L_o}$ \quad \small\text{total stretch,}	 & 
		$\lambda_e=\frac{L_e}{L_o}$ \quad \small\text{elastic stretch,}	\\[6pt]
		$\lambda_n=\frac{L_n}{L_o}$ \quad \small\text{permanent stretch,}		&
		$\lambda_c=\frac{L_c}{L_o}$ \quad \small\text{contour stretch,}
	\end{tabular}\end{equation}
with $L_o=b\, \sqrt{n_o}$ denoting the initial natural length. 

The natural stretch for the soft region can be deduced using Eqns.~\eqref{ln} and \eqref{eqn:stretch measures}:
\begin{equation} \label{eqn:lanb}	\lambda_n(\text{\small RH})=\frac{L_n(\text{\small RH})}{L_o}=
	\frac{\sqrt{n_o \  m(\text{\small RH})} \ b}{\sqrt{n_o \ m_o} \ b}=\sqrt{\frac{m(\text{\small RH})}{m_o}} \end{equation} with $m(\text{\small RH})/m_o$ given by Eqn.~\eqref{mmo} so that the variation of the natural length with humidity explicitly depends from physically based parameters.
On the other hand, the corresponding expression for the contour length is
$L_c={m\frac{n_o}{m}} \ b =n_o b,$
so that the contour stretch of the amorphous part is
\begin{equation} \label{eqn:lcs}
	\lambda_c=\frac{L_c}{L_o}=
	\frac{n_o \ b}{\sqrt{n_o \ m_o} \ b}=
	\sqrt{\frac{n_o}{m_o}}.
\end{equation}

	Under an additive assumption, the (Piola, engineering) stress is determined using Eqns.~\eqref{eqn:force-stretch}$_2$ and \eqref{eqn:stretch measures}
\begin{equation} \label{eqn:sigma stretch soft part}
	\sigma_{soft}=E \left[ 
	\left( \frac{\lambda_c-\hat{\lambda}_n(\text{\small RH})}{\lambda_c-\lambda}\right)^2-1
	\right], \end{equation}
where the permanent stretch depends by $\text{\small RH}$ and it is given by Eqn.~\eqref{eqn:lanb}  and Eqn.~\eqref{mmo}
whereas the contour stretch is constant (see Eqn.~\eqref{eqn:lcs}). Here
$E=N_a \kappa$ is the elastic modulus of the soft fraction with $N_a$ the number of chains per unitary reference area.
Indeed the total energy considering all the chains of the network can be determined as $\Phi=N_v\varphi_eL_c=N_aN_l\varphi_eL_c=N_a\varphi_e$ where for the sake of simplicity we consider $N_lL_c=1$, with $N_v,N_a,N_l$ the number of chains per unit volume, area, and length respectively.

\begin{figure}[t] \centering
			\includegraphics[width=\textwidth]{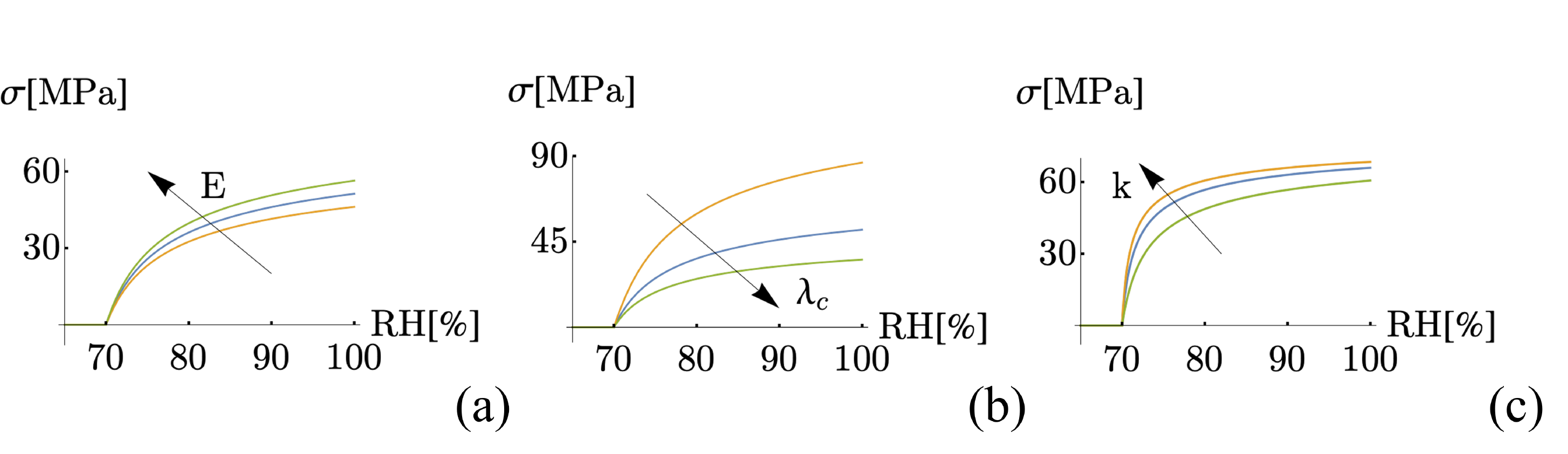}
\caption{\label{fig:mat_des} Influence of the material parameters on the macroscopic stress-stretch plot. 
		(a) Elastic modulus soft region $E=9,10,11$ MPa, $\lambda_c=1.5,\kappa=0.45$,
		(b) Contour stretch soft region $\lambda_c=1.35,1.5,1.65, E=10$ MPa, $\kappa=0.45$, 
		(c) constant depending on the H-bonds disruption kinetics $\kappa=1.1,2.2,3.3,\lambda_c=1.5, E=10$ MPa. $\text{\small RH}_c=70\%$. \label{fig:material_design}}
\end{figure}

Observe that the supercontraction stress $\sigma_{act}=\sigma_{soft}(1)$ depends on three only material parameters all having a clear microstructure interpretation. In Fig.~\ref{fig:material_design} we represent the possibility of designing the material response by changing the material parameters, namely the module of the soft fraction $E$, the contour stretch of the chains of the soft fraction $\lambda_c$ and the constant  $k$ regulating on the H-Bonds disruption kinetics. The influence of possible deformable constraints are considered in the following. In the simulations reported in Fig.~\ref{fig:material_design} typical values of spider silks have been used for material parameters, as it will be evident from the experimental validation in the following section. For the elastic modulus and the contour stretch, variations of $\pm$10\% were considered, while for the constant $k$ a variation of $\pm$50\% was considered to obtain an appreciable variation in the stress-stretch plot.

\section{Experimental validation}
\begin{figure}[h!] \centering 	
	\includegraphics[width=.6\textwidth]{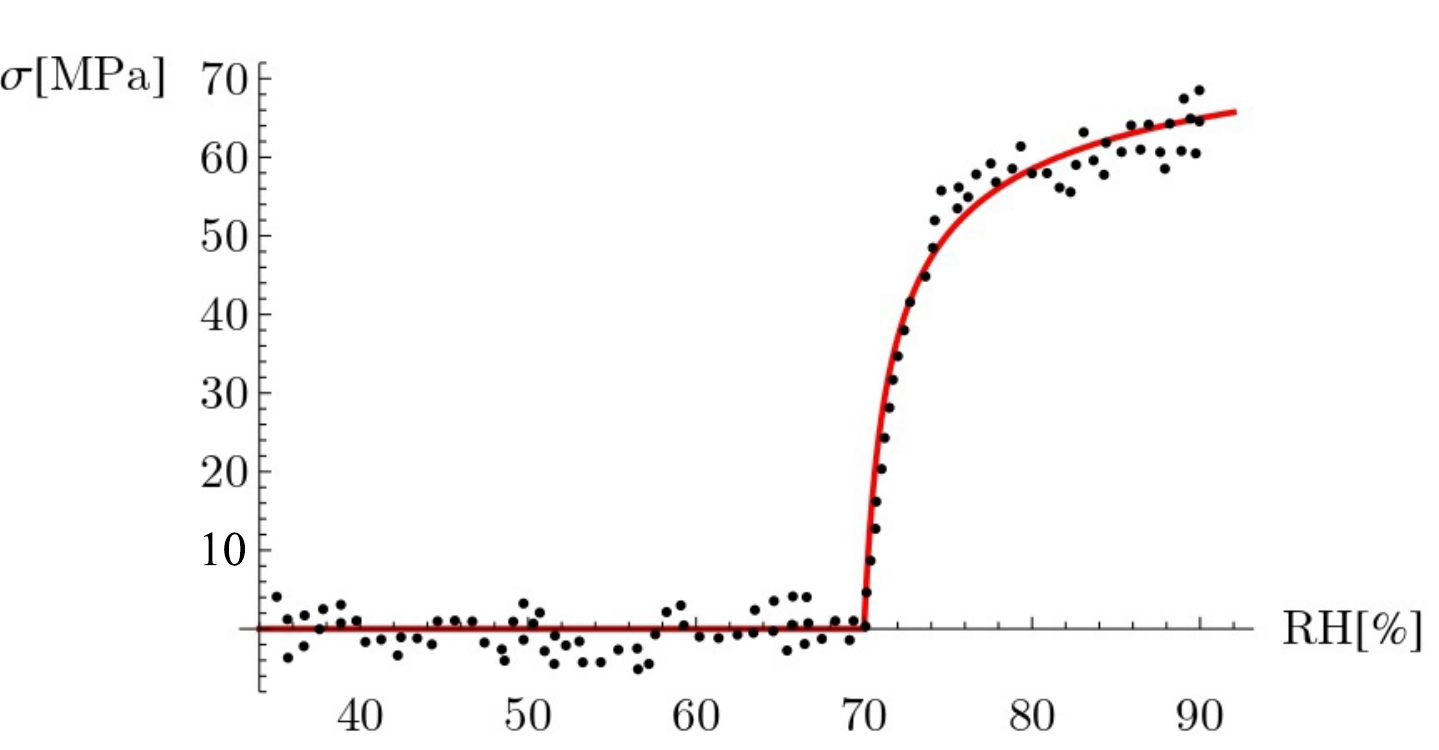} \vspace{-.3cm}
	\caption{Stress-humidity curve for an {\it Argiope trifasciata} spider: dots are the experiments reproduced from \cite{Guinea2003}, whereas the continuous line represents the theoretical prediction with parameters $E=10.3$ MPa$,\lambda_c=1.5,\kappa=2.22,\text{\small RH}_c=70\%$. } \label{fig:scs}  \end{figure}

To test the effectiveness of the proposed model in quantitatively describing the experimental behavior, we consider in Fig.~\ref{fig:scs} a test on spider dragline silk ({\it Argiope trifasciata}) reproduced from \cite{Guinea2003}, in an experiment when the fiber is clamped at maximum length with zero force and exposed to monotonically increasing humidity content while measuring the force. In the figure the value of the engineering stress (Piola Kirchhoff stress, force divided by initial cross-sectional area) is reported.

As expected for restrained fibers tested in controlled environment with monotonically increasing {\small RH} a sudden development of supercontraction stress occurs only once the {\small RH} overcomes the threshold of $\text{\small RH}_c\cong70\%$ with mainly no stress variation below this critical threshold. A steep increasing of stress is observed in the narrow range between {\small RH}$\sim70\%$ and  {\small RH}$\sim75\%$. For higher {\small RH}s the stress keeps growing, but with a much lower slope until {\small RH}=100$\%$ when the maximum actuation stress is attained.

Notice that the optimal fit parameters are fully compatible with the ones employed in \cite{FDPP} to reproduce the stress-stretch behavior of fibers of the same spider species ({\it Argiope trifasciata})  where they were deduced based on their micromechanical interpretation. Indeed here we assume $E=10.3$ MPa$,\lambda_c=1.5$ for the elastic modulus and the contour stretch, respectively, in place of $E=13.5$ MPa$,\lambda_c=1.62$ employed in \cite{FDPP} for the soft region. Likewise, the assumed supercontraction threshold $\text{\small RH}_c=70\%$ match the one experimentally found during the experimental test of the fiber \cite{Guinea2003}. These differences are then justified by the observation that the experiment refers to a different tested silk of the same spider species. The only ``arbitrary fitting" parameter is therefore the constant $k$ regulating the decreasing of the number of H-bonds when the humidity overcomes the critical threshold. To the knowledge of the authors, no experimental values are available so that we assume $k\cong 2$ as a best fit parameter. 
The model turns out to be accurate in quantitatively reproducing the experimental behavior, based on the obtained final relation \eqref{eqn:sigma stretch soft part}.

\section{Elastic interaction}
\begin{figure}[h] \centering 	
	\includegraphics[width=1 \textwidth]{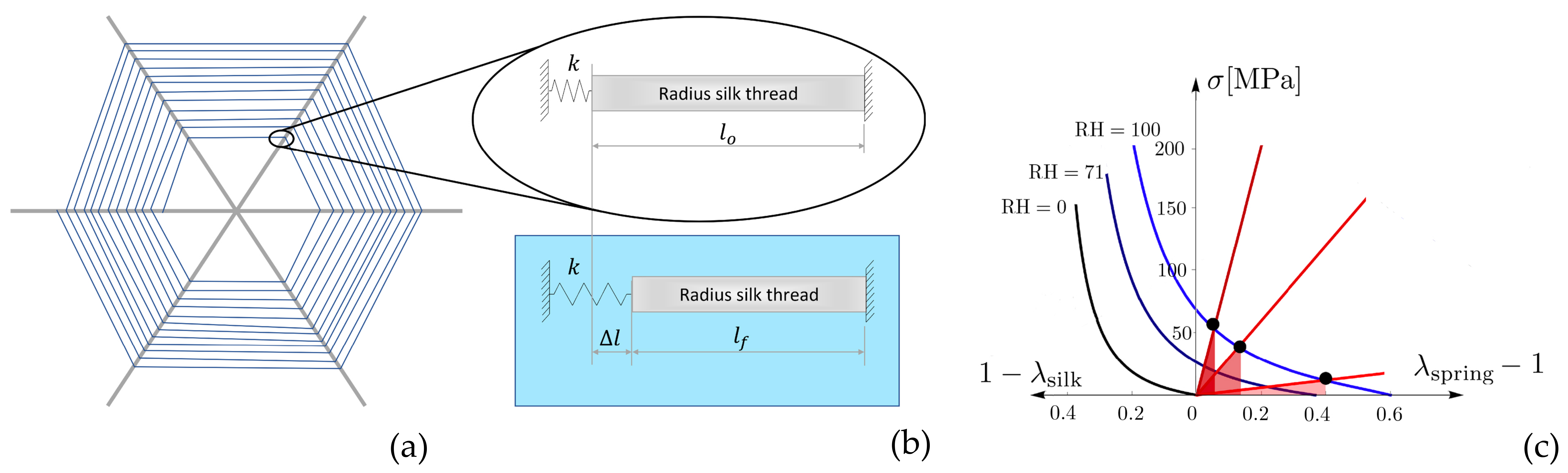}
	\caption{\label{fig:sch1} (a) scheme of a spider web and the radius silk thread-spring interaction. The spider web is composed by radii (grey) and capture spiral (blue). Due to different composition, the radii may undergo contraction in humid conditions, whereas the capture spiral threads do not contract. (b) simple scheme of the elastic interaction between the radius silk contracting and elongating spiral silks modeled as elastic springs. (c) Stress-strain behaviors of the silks at different humidity and different stiffnesses of the springs represented in the same diagram using the compatibility condition \eqref{comp}. Different equilibrium configuration corresponding to humidity saturation and different actuated device stiffness are represented by black dots. Assumed parameters: $E=10$ MPa, $\lambda_c=1.5,$ $\kappa=2.22,\text{\small RH}_c=70\%, \text{\small RH}=100\%,\mu=1$ MPa, $k=30,300,1000$ MPa}\end{figure}

We previously introduced two special cases of boundary conditions, namely the case of a restrained thread with fixed end-to-end distance studied in the previous section, and the case of unrestrained supercontraction discussed in detail in \cite{FDPP}. In this section we explore the actuation properties of the spider silk thread during supercontraction in the more realistic case when elastic interactions at the boundary are considered, as schematized in Fig.~\ref{fig:sch1}. This scheme mimics for example the case of spider webs where the different threads, {\it i.e.} the radii and the capture spiral, respond (and thus contract) in different ways to changes in humidity environment. In particular, the capture spiral is composed of elastic flagelliform silk that is coated by hygroscopic glue droplets. The glue droplets contain water that maintains flagelliform silk in a continuously supercontracted state \cite{Guinea2010recovery} so that no further contraction occurs in wet external environment. 
On the other hand, the Major Ampullate (MA) silk forming the radii contracts when the threads are immersed in wet environment. In the simplified scheme for studying the elastic interaction among the threads, the MA silk is represented by the radius silk thread that when contracts may elongate the spring (capture spiral thread). In the following, for simplicity, we consider the capture spiral thread elasticity by constraining the radial thread with a linear spring.  
 


Since we are interested to work and power density, hereon we can assume unit reference area and length for both the actuated device and spider thread.
We have then to respect the simple equilibrium condition (suppose for simplicity of notation that the two devices have unitary area)
\begin{equation}  \label{eqn:eql} \sigma_{\text{spring}}=\sigma_{\text{silk}}.\end{equation}
On the other hand the compatibility condition of the strain, imposing that the shortening of the silk thread from the dry condition, measured by $1-\lambda_{\text{silk}}$, equals the elongation of the activated device 
$\lambda_{\text{spring}}-1$, gives 
\begin{equation}  \label{comp}\lambda_{\text{spring}}-1=1-\lambda_{\text{silk}}\end{equation}  
with
\begin{equation} \sigma_{\text{spring}}=k_{\text{spring}}(1-\lambda_{\text{spring}}). \end{equation}

Then the equilibrium stretch $\lambda_{\text{eq}}$ is uniquely defined using Eqn.~\eqref{eqn:eql} by solving
\begin{equation} \label{eqn:sigma stretch silk spring}
	E\left[ 
	\left( \frac{\lambda_c-\hat{\lambda}_n(\text{\small RH})}{\lambda_c-\lambda_{\text{eq}}}\right)^2-1
	\right]+\mu \left( \lambda_{\text{eq}}- \frac{1}{\lambda_{\text{eq}}^2}\right)=k_{\text{spring}}(1-\lambda_{\text{eq}}).\end{equation}
 
\begin{figure}[h] \centering
{\includegraphics[width=0.76\textwidth]{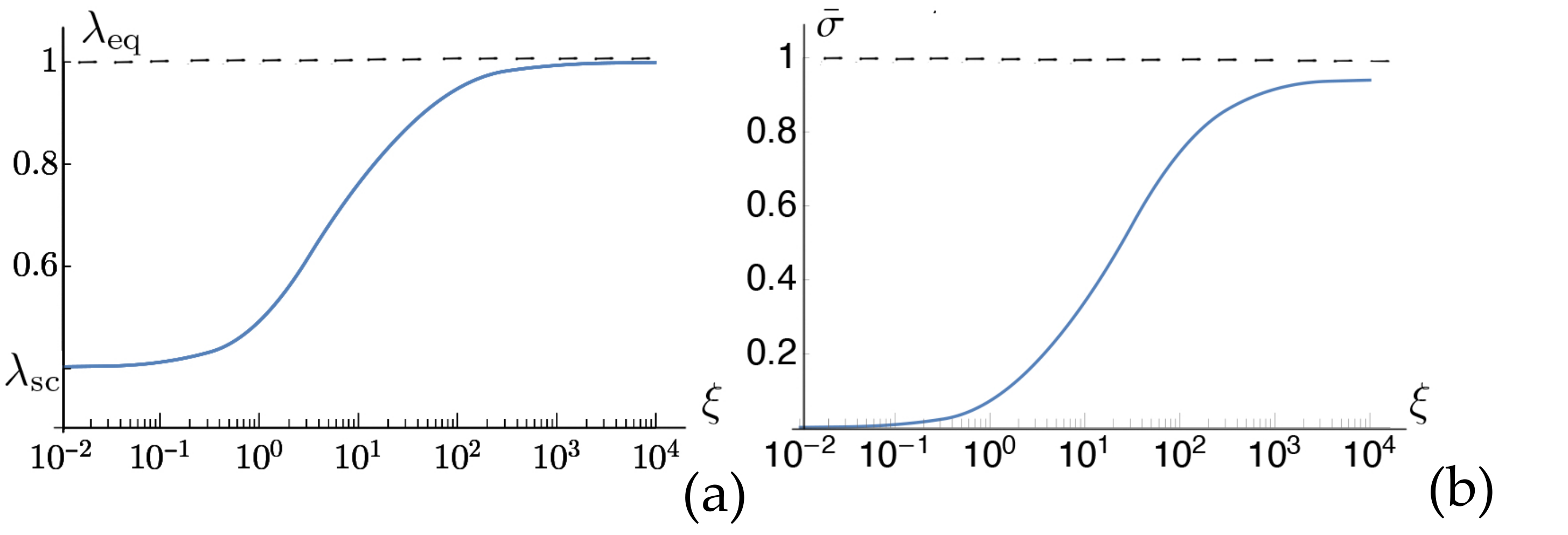}}
\caption{\label{fig:lcs_Espring}Actuation dependence on spring modulus. Assumed parameters: $E=10$ MPa, $\lambda_c=1.5,$ $\kappa=2.22,\text{\small RH}_c=70\%, \text{\small RH}=100\%,\mu=1$ MPa}
\end{figure}

By increasing humidity, as supercontraction occurs, the equilibrium stretch $\lambda_{\text{eq}}$ grows as shown in Fig.~\ref{fig:sch1} (c). In the figure, using Eq.~\eqref{comp} we represent the stress of the silk thread and actuated device in the same diagram, so that the equilibrium configurations correspond to the intersections of the two stress-strain curves. The corresponding actuation work is represented by shaded areas.
 The dependence of the stretch and non-dimensional stress $\bar \sigma:=\frac{\sigma_{\text{act}}}{\sigma_{\text{act,max}}}$ on the stiffness ratio $\xi=\frac{k_{\text{spring}}}{E}$  is represented in Fig.~\ref{fig:lcs_Espring}.
Notice that for high values of $\xi$ the stretch reaches the saturation value corresponding to the restrained supercontraction ($\lambda=1$) and the silk approaches the restrained condition actuation stress ($\sigma_{\text{act}}=\sigma_{\text{act,max}}$). On the other hand, the supercontraction stress is zero for low values of $\xi$ (corresponding to unrestrained silk thread), where the contraction stretch attains its minimum value $\lambda_{\text{eq}}=\lambda_{\text{sc}}$ known as supercontraction stretch.

\section{Water to the ropes!}
In 1586 Pope Sisto V, wanting to embellish Saint Peter’s Square (Rome, Italy), ordered that the large obelisk -that is still admired there- be erected.
The work, which was entrusted to the architect Domenico Fontana, presented serious difficulties. The obelisk weighed 350 tons and was 25 meters high, so that Fontana had to do calculations and engage scaffolding, winches, pulleys, hemp ropes, 800 men, and 140 horses.
The obelisk was almost in place when people saw the ropes overheat dangerously, with the risk that they caught fire. The monolith would have fallen to the ground. Then in the silence there was a scream: ``Daghe l'aiga ae corde!'' (expression of the Ligurian language meaning ``Water to the ropes").
The advice was immediately followed with excellent results. To thwart the danger had been Captain Benedetto Bresca, Ligurian sailor of Sanremo, who knew well that the hemp ropes are heated by the clutch of the winches and also shorten when they are wet. Thus the water-induced contraction of hemp ropes was employed to erect the obelisk thus using it as a humidity driven actuator. 
\begin{figure}[h] \centering 		\includegraphics[width=\textwidth]{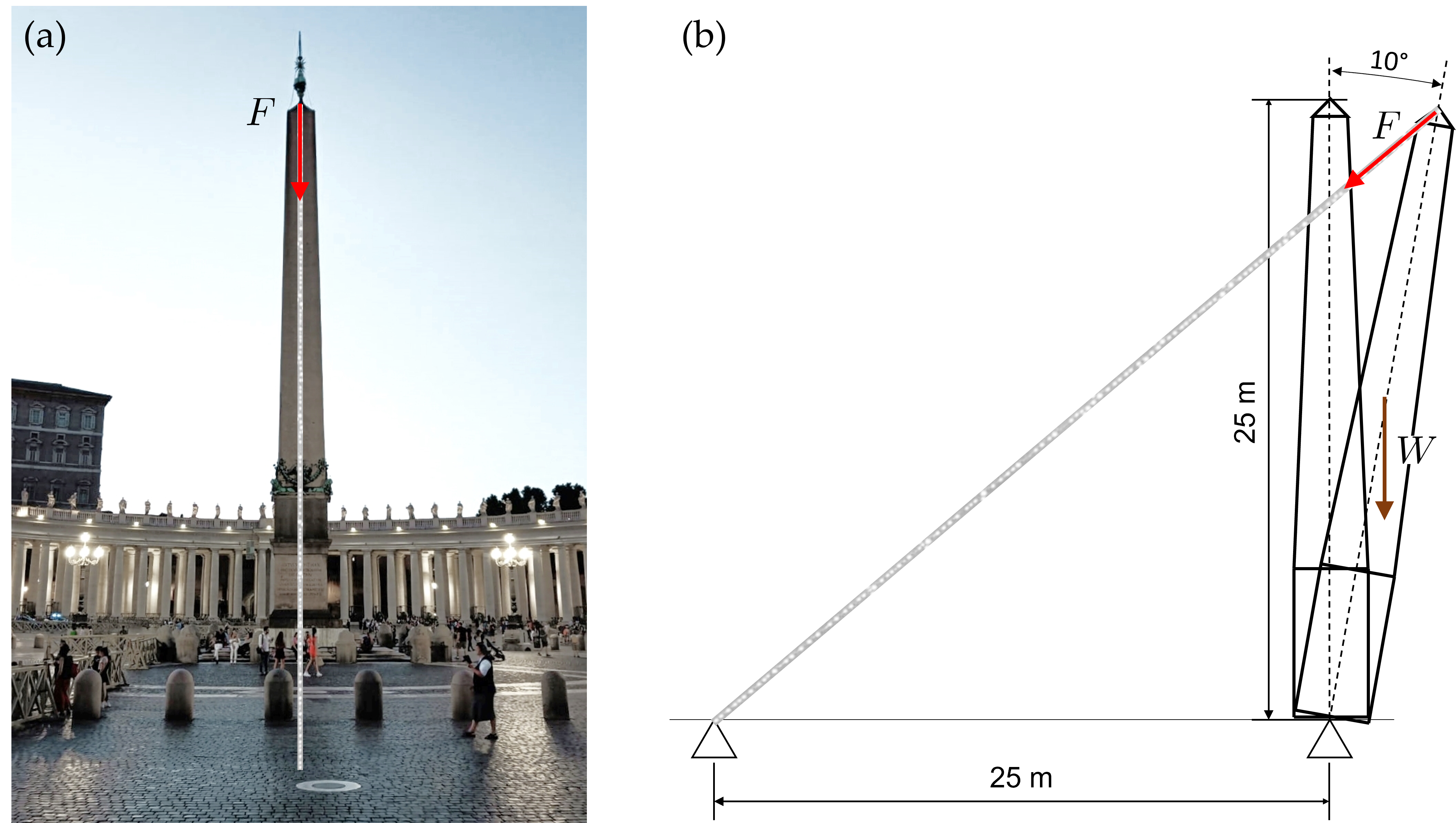}
	\caption{\label{fig:obelisco_vaticano} Scheme of the equilibrium recovery for the obelisk, considering a weight of 3.5~MN applied in the center of mass in a configuration of 10$^{\circ}$ from the equilibrium vertical configuration, obtained based on supercontraction spider silk actuation. Considering the parameters obtained for the {\it Argiope trifasciata} silk fiber (Fig.~\ref{fig:scs}), we calculated the diameter of the spider silk thread needed that is only $10$ cm.}\end{figure}

Here, to exemplify through a `real' example the strong actuation properties of spider silks when subjected to humidity,  we consider the possibility of erecting the obelisk by employing the supercontraction properties of the spider silk. More explicitly, we determine the diameter of the silk thread needed to rotate the obelisk in the vertical direction, starting from a misalignment angle of 10$^\circ$ and assuming that the distance of the constraint is equal to the height (25 m) of the obelisk (see Fig.~\ref{fig:obelisco_vaticano}). For simplicity we consider the center of gravity at the half of the height  where the 350 tons weight is applied. 

First observe that according with the description of previous section, also in this case the length of the thread is not fixed, because it decreases as the obelisk rotates. In particular, by simple geometric argument it is easy to determine that to recover the vertical configuration the thread should undergo a contraction $\lambda_{eq}=0.92$. Based on the previously considered physical parameters (see Fig.~\ref{fig:lcs_Espring}(a)) such a contraction corresponds to consider an elastic spring with a stiffness ratio $\xi=60$. We may then determine (see Fig.~\ref{fig:lcs_Espring}(b)) the related actuation stress $\sigma_{act}=47$ MPa. The diameter of the spider silk thread  is then calculated by a simple rotational equilibrium equation (see Fig.~\ref{fig:obelisco_vaticano}(b)) and it results of about $10$~cm. 

\section{Actuation properties}
\begin{figure}[htb] \centering
				{\includegraphics[width=\textwidth]{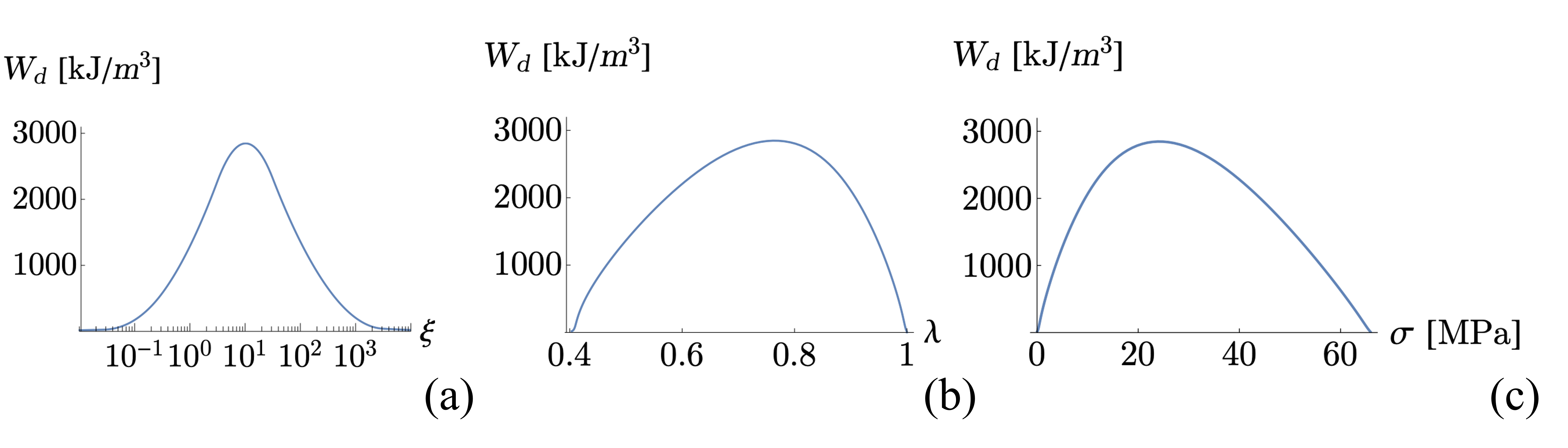}}	\caption{\label{fig:workdensity} Work density generated by spider dragline silk as a function of the stiffness ratio~(a), supercontraction stretch (b) and stress (c). Assumed parameters: $E=10$ MPa, $\lambda_c=1.5,$ $\kappa=2.22,\text{\small RH}_c=70\%, \text{\small RH}=100\%,\mu=1$ MPa}
\end{figure}

To exemplify the incredible actuation properties of the considered spider silk, we may evaluate the work density of previous example as 
$ W_d=\frac{1}{2}\sigma_{act}(1-\lambda_{\text{spring}})=1814\,\frac{\text{kJ}}{\text{m}^3}.	$
The work density dependence on the boundary conditions is described in Fig.~\ref{fig:workdensity}, where it is represented as a function of the stiffness ratio, supercontraction stretch, and stress. The maximum attainable work density $W_d=2846\frac{\text{kJ}}{\text{m}^3}$, considering a spider silk density of $\rho=1.3$ g/cm$^3$ \cite{Stautter_1994}, corresponds to a work capacity of $W_c=2.19$ kJ/kg. To the knowledge of the authors, this value is higher than the maximum ever recorded for a moisture powered actuator, namely the hybrid poly(diallyldimethylammonium chloride) Carbon nanotube yarns (PDDA/CNT) actuator providing a work capacity up to $W_c=2.17$ kJ/kg \cite{Kim2016}.
These values are well above (over 50 times) the mean work capacity of human muscle of $W_c=0.039$ kJ/kg \cite{madden2004artificial}.

Eventually, by considering a realistic  contraction time of $t=3$~s \cite{agnarsson2009spider} we obtain an actuation power density
\begin{equation}
	P=\frac{W_d}{\rho \, t}=730\,\frac{\text{W}}{\text{kg}}.
\end{equation}
Also this value is of absolute relevance compared with human muscle ($P=50\div$280 W/kg) and Carbon Nanotube actuators ($P=10\div$270 W/kg) \cite{madden2004artificial}.

Further, we represent the operating points in the stress-stretch plot as intersections among the stress-stretch curve of the silk threads and some possible curves of the actuated device, in order to point out some interesting considerations on the actuation properties of the spider silk. In particular, in Fig.~\ref{fig:sch1}(c) we show the case of linear actuated device, for different {\small RH} and for different spring stiffness.
In the case of the highest {\small RH} (blue curve), we indicate the different operating points (as black circles) and the corresponding work (as red areas below the spring curves). Note that the maximum area is obtained at an `intermediate' value of the spring modulus, whereas for the highest and lowest spring modulus, the work decreases and it is null with spring modulus ideally going to zero and infinity, corresponding to the special cases of unrestrained and full restrained supercontraction,  respectively, typically considered in the experiments. 
These observations are consistent with those on the work density in Fig.~\ref{fig:workdensity}(a) and remark the crucial role of actuated device stiffness in defining the optimal actuation properties of the system.

\subsection{Bi-stable actuated device}
\begin{figure}[b!] \centering
			{\includegraphics[width=.95 \textwidth]{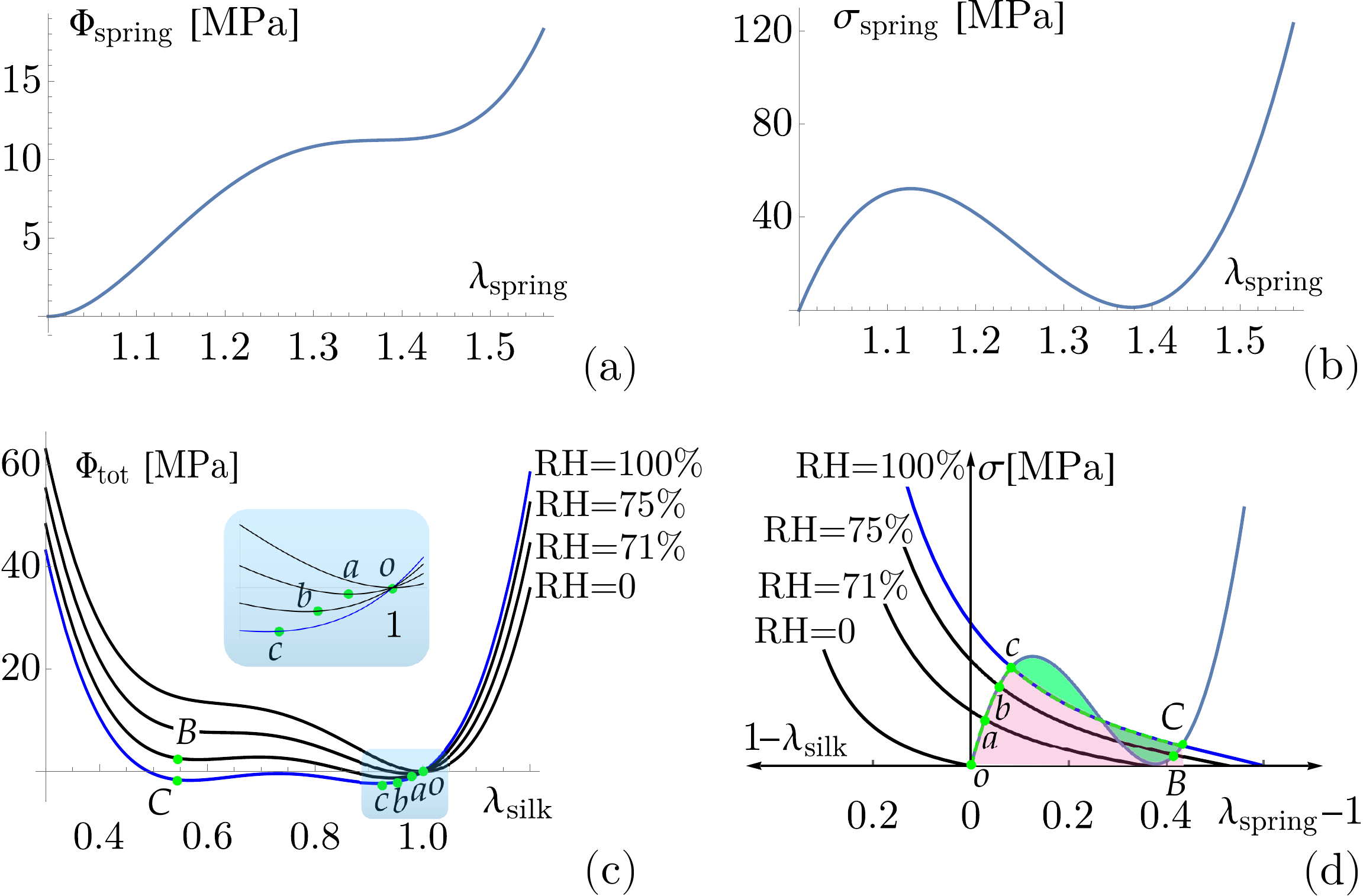}}	\caption{\label{fig:actuation}
	Non-convex energy-strain a) and non-monotonic nonlinear relation b) of the actuated device. Dots represents stable equilibrium solutions for different values of  $ \text{\small RH}$ in the $\Phi_{\text{tot}}-\lambda_{\text{silk}}$ space c) and stress-strain space d). Stress-strain behaviors of the silks at different humidity and different stiffnesses of the springs represented in the same diagram using the compatibility condition \eqref{comp}. Assumed values: nonlinear spring
	$\sigma_{\text{spring}}(\lambda_{\text{spring}})=6.5 \times 10^9(\lambda_{\text{spring}}(4.653 + (\lambda_{\text{spring}}-3.755)\lambda_{\text{spring}})-1.898)$; silk, $\lambda_c=1.5,$ $\kappa=2.22,\text{\small RH}_c=70\%, \text{\small RH}=0,71,75,100\%,\mu=1$ MPa.}
\end{figure}

Eventually we show the possibility of obtaining an even significantly higher work by considering non linear effects. In particular we investigate the interesting case of bistable actuated devices with a region of material instability where the energy is concave and in particular we consider the simple case of a cubic stress-strain relation $\sigma_{spring}(\lambda)$ with a negatively sloped strain domain (see Fig.~\ref{fig:actuation}(b)). Thus if we write the non-convex energy of the nonlinear actuated device (see Fig.~\ref{fig:actuation}(a))
\begin{equation}\Phi_{\text{spring}}(\lambda_{\text{spring}})=\int_{1}^{\lambda_{\text{spring}}}\sigma_{\text{spring}}(\lambda)\, d\lambda\end{equation}
and the silk energy density 
\begin{equation}\Phi_{\text{silk}}(\lambda_{\text{silk}})=\int_{1}^{\lambda_{\text{silk}}}\sigma_{\text{silk}}(\lambda)\, d\lambda\end{equation}
with total energy (assume for simplicity of notation unitary length)
\begin{equation}\Phi_{\text{tot}}(\lambda_{\text{silk}})=\Phi_{\text{spring}}(\lambda_{\text{spring}})+\Phi_{\text{silk}}(\lambda_{\text{silk}})=\Phi_{\text{spring}}(\lambda_{\text{spring}})+\Phi_{\text{silk}}(2-\lambda_{\text{spring}}),\label{phitot}
\end{equation}
the stationarity equation \begin{equation}\frac{d \Phi_{\text{tot}}(\lambda)}{d \lambda}=0\end{equation} delivers again the equilibrium condition \eqref{eqn:eql}. This equation (see Fig.~\ref{fig:actuation}(d)) has one single solution for low values of $\text{\small RH}$ belonging to the right stable equilibrium branch. When $\text{\small RH}$ is increased the system has two distinct equilibrium solutions. For example in the figure for $\text{\small RH}=75 \%$ we denote the solutions by $b$ (first stable branch) and $B$ (second stable branch) with
\begin{equation}\begin{array}{lll}\Phi_{\text{tot}}(\mbox{B})-\Phi_{{\text{tot}}}(\mbox{b})&=&\displaystyle \int_{\lambda_b}^{\lambda_B}\Phi'_{\text{tot}}(\lambda)\, d\lambda\vspace{0.2 cm}\\
&=&\displaystyle \int_{\lambda_b}^{\lambda_B}\left(\sigma_{\text{spring}}(\lambda_{\text{spring}})-\sigma_{\text{silk}}(2-\lambda_{\text{spring}})\right) d\lambda_{\text{spring}}\end{array}.\end{equation}

We are then lead to the analogue of the Maxwell construction for bistable devices (see {\it{e.g.}} \cite{PUGLISI20001}) and the system has a strain discontinuity when the area between the actuated and activation stress-strain systems equals zero. In Fig.~\ref{fig:actuation}(d) we optimized such condition by choosing the material response of the non-linear spring in such a way that the Maxwell condition is attained exactly in correspondence with the fully wet condition $\text{\small RH}=100\%$ (see the shaded green area in Fig.~\ref{fig:actuation}(d) and the corresponding energy in Fig.~\ref{fig:actuation}(c)). Thus if we suppose to increase the humidity starting from the dry condition (point $o$ in the figure) the system has one single equilibrium configuration up to  $\text{\small RH}\sim 70\%$ (see point $a$ in the figure). For larger values of  $\text{\small RH}$ two distinct solutions are possible (the intersections with the negatively sloped part of the sigmoidal curve are unstable equilibrium solutions \cite{PUGLISI20001}), but only when the equal area condition is attained the solution in the larger strain stable equilibrium branch represents the lowest energy. The system then follows the path $o-c-C$ when it `jumps' to the second branch and the Maxwell condition is attained. As a result the system exhibits a large increase of actuation work (shaded purple area in the figure) that for the same spider silk properties considered above leads to the very large work density
$W_d=10.9 \frac{\text{kJ}}{\text{m}^3}$.

\section{Conclusions}
We addressed the issue of quantifying the supercontraction stress arising when a restrained spider silk fiber is hydrated at different humidity conditions and the corresponding actuation properties.
By extending the recent approach that we proposed in \cite{FDPP}, we deduced a microstructure inspired model taking into account of the H-bond disruption process induced by external humidity in the hydrophilic macromolecular chains composing the spider silk. 
The number of H-bonds determines then the natural (zero-force) length of a chain that can be quantified by means of classical statistical mechanics results. In particular, the force-elongation relationship is obtained by considering a WLC type energy with a humidity-dependent natural configuration. We then deduce the macroscopic behavior based on the classical affinity hypothesis identifying the macroscopic stretches with the macromolecular ones.
The model has been tested against experiments of restrained silk fibers  exposed to increasing moisture content while
measuring the fiber stress. By considering material parameters of the thread coherent with previously determined ones \cite{FDPP}, we obtain that the theoretical prediction well reproduces the experimental behavior. We then exploit the important influence of boundary conditions in determining the maximum supercontraction stress and stretch, neglected by previous literature.

We then apply the proposed model, by considering known experimental values of the material parameters, to predict the supercontraction stress arising in spider threads at different humidities and boundary conditions reproducing effective real biological devices \cite{greco2021tyrosine}. In particular we stress the fundamental role of the stiffness of the constraint device, representing the other spider threads in the web or the possible device considered in the actuation process. This would be clearly of interests in the field of the mechanical actuation by using the humidity as driving load \cite{dong2021programmable}.

Moreover we determine, based on the introduced elastic interaction hypothesis,  the important power and work densities parameters, showing the superior properties of humidity based actuators constituted by spider silks as compared with the most performing artificial materials.
Eventually we show the possibility of obtaining even higher work by considering multistable responses of the actuated system. 

\section*{Declaration of Competing Interest}
The authors declare that they have no known competing financial interests or personal relationships that could have appeared to influence the work reported in this paper.	

\section*{Acknowledgments}
Funding: GP has been supported by the Italian Ministry MIUR-PRIN project 2017KL4EF3 and by GNFM (INdAM) and NMP by the European Commission under the FET Open “Boheme” grant no. 863179 and by the Italian Ministry of Education MIUR under the PON ARS01-01384- PROSCAN Grant and the PRIN-20177TTP3S.

\bibliographystyle{unsrt}
\bibliography{scs.bib}

\end{document}